\documentclass[twocolumn,english,english,preprintnumbers,prb,superscriptaddress]{revtex4}
\usepackage[T1]{fontenc}
\usepackage[latin9]{inputenc}
\usepackage{amsmath}
\usepackage{amssymb}
\usepackage{esint}
\usepackage[normalem]{ulem}
\usepackage{graphicx}
\usepackage{graphics}
\usepackage{epstopdf}
\usepackage{epsfig}
\usepackage{amsfonts}
\usepackage{bm}
\usepackage{amscd}
\usepackage{babel}
\usepackage{babel}
\usepackage{babel}
\usepackage{babel}
\usepackage{babel}

\makeatletter
\@ifundefined{textcolor}{}
{
 \definecolor{BLACK}{gray}{0}
 \definecolor{WHITE}{gray}{1}
 \definecolor{RED}{rgb}{1,0,0}
 \definecolor{GREEN}{rgb}{0,1,0}
 \definecolor{BLUE}{rgb}{0,0,1}
 \definecolor{CYAN}{cmyk}{1,0,0,0}
 \definecolor{MAGENTA}{cmyk}{0,1,0,0}
 \definecolor{YELLOW}{cmyk}{0,0,1,0}
 }
\@ifundefined{definecolor}{\usepackage{color}}{}
\input{tcilatex}
\makeatother

\begin{document}

\title{Non-adiabatic Hall effect at Berry curvature hot spot}


\author{Matisse Wei-Yuan Tu}
\email{kerustemiro@gmail.com}
\affiliation{Department of Physics, The University of Hong Kong, and HKU-UCAS Joint Institute of Theoretical and Computational Physics at Hong Kong, China}
\author{Ci Li}
\affiliation{Department of Physics, The University of Hong Kong, and HKU-UCAS Joint Institute of Theoretical and Computational Physics at Hong Kong, China}
\author{Hongyi Yu}
\affiliation{Guangdong Provincial Key Laboratory of Quantum Metrology and Sensing and School of Physics and Astronomy, Sun Yat-Sen University (Zhuhai Campus), Zhuhai 519082, China}
\affiliation{Department of Physics, The University of Hong Kong, and HKU-UCAS Joint Institute of Theoretical and Computational Physics at Hong Kong, China}
\author{Wang Yao}
\email{wangyao@hku.hk}
\affiliation{Department of Physics, The University of Hong Kong, and HKU-UCAS Joint Institute of Theoretical and Computational Physics at Hong Kong, China}

\begin{abstract}
Hot spot of Berry curvature is usually found at Bloch band anti-crossings, where the Hall effect due to the Berry phase can be most pronounced. With small gaps there, the adiabatic limit for the existing formulations of Hall current can be exceeded in a moderate electric field. Here we present a theory of non-adiabatic Hall effect, capturing non-perturbatively the across gap electron-hole excitations by the electric field. We find a general connection between the field induced electron-hole coherence and intrinsic Hall velocity. In coherent evolution, the electron-hole coherence can manifest as a sizeable ac Hall velocity. When environmental noise is taken into account, its joint action with the electric field favors a form of electron-hole coherence that is function of wavevector and field only, leading to a dc nonlinear Hall effect. The Hall current has all odd order terms in field, and still retains the intrinsic role of the Berry curvature. The quantitative demonstration uses the example of gapped Dirac cones, and our theory can be used to describe the bulk pseudospin Hall current in insulators with gapped edge such as graphene and 2D MnBi$_{2}$Te$_{4}$.
\end{abstract}

\maketitle

\section{Introduction}

Intrinsic Hall current arising from Berry phase effect of Bloch electrons is
of long standing interest.~\cite{Nagaosa101539,Xiao101959} The Hall conductivity is
determined by the Berry curvature, which, being inversely proportional to
the square of the band separation, is most pronounced at band anti-crossings
with small gap. A seminar example is the gapped Dirac cones found in
graphene with inversion symmetry breaking,~\cite%
{Xiao07236809,Hunt131427,Woods14451,Gorbachev14448,Sui151027,Shimazaki151032,Lensky15256601,Song1510879}
surface bands of ultrathin films of topological insulators~\cite{Zhang10584,Lu10115407,Liu12036805}
and magnetic topological insulators such as MnBi$_2$Te$_4$,~\cite{Zhang19206401,Li19eaaw5685,Gong19076801,Otrokov19416,Liu2001.08401,Deng1904.11468v1,Liu2005733,Ge1907.09947}
which are all drawing great interest. Berry curvature distribution features
hot spot at the Dirac point, where a small momentum space area can enclose a
sizeable flux, underlying a strong Hall response.

Existing formulations of Hall effects are limited to the adiabatic regime,
i.e. assuming a sufficiently large gap preventing electric field to create
interband excitations.~\cite{Xiao101959} The first order adiabatic
approximation gives the linear response, with an intrinsic Hall conductivity
equal to the Berry curvature flux enclosed by the equilibrium Fermi sea.~\cite%
{Thouless82405} Generalization of the Hall effect to second order in
electric field has also been progressed along two lines, both in the
adiabatic limit,~\cite%
{Gao14166601,Deyo091917v1,Moore10026805,Sodemann15216806,Du193047,Nandy19195117,Xiao19165422}
by either including a field correction to the Berry curvature,~\cite%
{Gao14166601} or the extrinsic mechanism of disorder scatterings and
nonequilibrium carrier distribution.~\cite{Deyo091917v1,Moore10026805,Sodemann15216806,Du193047,Nandy19195117,Xiao19165422} 
Experimental observations of second order Hall effect have been reported in
2D WTe$_2$,~\cite{Xu18900,Ma19337,Kang19324}
where the Fermi energy also lies near a Berry curvature hot spot from a band
anti-crossing.~\cite{Xu18900,Ma19337} These findings have stimulated remarkable
interests on nonlinear Hall responses.

A Berry curvature hot spot, however, is always accompanied by a small gap.
In the aforementioned materials, the gap is within a few tens of meV,~\cite%
{Hunt131427,Woods14451,Gorbachev14448,Zhang19206401,Liu2005733,Xu18900,Ma19337,Qian141344}
which is not sufficient to guarantee adiabaticity in experimental
conditions, especially when a large current is desired. In addressing the
strong Hall effect at such Berry curvature hot spots, going beyond the
adiabatic regime is therefore highly relevant, which can in principle lead
to non-perturbative response to the electric field.

It is also known that, when Fermi energy lies in the gap, the Dirac cone has
a half-quantized Hall conductivity in the linear response, $\frac{1}{2}
\text{sgn}(\Delta) \frac{e^2}{h}$, where $\Delta$ is the Dirac mass (gap).~\cite{Qi08195424,Yao09096801}
In reality, Dirac cones come in pairs. For example, graphene has a pair of
cones with opposite (same) $\Delta$ forming the $K$ and $-K$ valleys at
Brillouin zone corners,~\cite{Xiao07236809} when the gap is opened by
inversion (time reversal) symmetry breaking.~\cite{Xiao07236809,Haldane882015}
And the recently discovered even-layer MnBi$_2$Te$_4$ ultrathin films
feature two cones with opposite (same) $\Delta$ on the top and bottom
surfaces respectively, when the magnetic order is layer anti-ferromagnetic
(ferromagnetic).\cite{Liu2005733}
Two cones of same $\Delta$ makes a quantum anomalous Hall insulator, where
the bulk-edge correspondence dictates a chiral edge channel inside the bulk
gap. Opposite $\Delta$ for the pair of cones, labeled by the valley or layer
pseudospin, means a half-quantized pseudospin Hall conductance in the gap.
In such case, however, the edge can also remain gapped as in the case of
graphene and MnBi$_2$Te$_4$,~\cite{Yao09096801,Li19eaaw5685}
and the absence of conduction channels at Fermi energy raises the intriguing
issue of how the valley or pseudospin current is sustained.~\cite%
{Gorbachev14448,Sui151027,Shimazaki151032}

Here we present a theory of non-adiabatic Hall effect, capturing the across
gap excitations by the electric field in the non-perturbation regime. In the
coherent evolution in constant electric field, a massive Dirac electron is
shown to develop a sizeable \textit{ac} Hall velocity, lying in the field
induced interband coherence dependent on evolution history. When
environmental noise is taken into account, the joint action of electric
field and decoherence favors certain form of interband coherence that is
function of wavevector and field only, with memory effect erased. The
corresponding Hall velocity is given by the Berry curvature times a
normalization factor non-perturbative in field. This leads to a $\mathit{dc}$
nonlinear Hall current, where the intrinsic contribution contains all odd
order terms in field. A general connection between field induced interband
(electron-hole) coherence and intrinsic Hall effect is thus established in
both the coherent and incoherent dynamics, with the known linear response
result properly reproduced. For insulator with gapped edge like the graphene
and MnBi$_2$Te$_4$ examples, the pseudospin Hall current can be sustained in
the bulk as the interference of the field induced electron-hole pair
excitations with the Fermi sea background. Our result is applicable to a
general band anti-crossing with a narrow gap.

\section{Results} 
In a homogeneous electric field $\boldsymbol{E}$, the
time evolution of an electron is described by
\begin{equation}
\mathcal{H}\left(\boldsymbol{k}\right)\left\vert u\right\rangle
=i\hbar\left\vert \dot{u}\right\rangle ,  \label{TDSE-1}
\end{equation}
where $\hbar \frac{d}{dt} \boldsymbol{k}= - e\boldsymbol{E}$. The
instantaneous eigenstates of $\mathcal{H}\left(\boldsymbol{k}\right)$ are
the Bloch functions: $\mathcal{H}\left(\boldsymbol{k}\right)\left\vert{u}_{n,%
\boldsymbol{k} }\right\rangle= \varepsilon_{n,\boldsymbol{k}} \left\vert{u}%
_{n,\boldsymbol{k}}\right\rangle$, $\varepsilon_{n,\boldsymbol{k}}$ giving
the band dispersion. For electron initially in a band $n$, the zeroth order
adiabatic evolution is given by $\left\vert{u}\right\rangle\approx e^{i
d_{n}\left(t\right) } e^{i\gamma_{n} (\boldsymbol{k})} \left\vert{u}_{n,%
\boldsymbol{k}}\right\rangle$. $d_{n}$ is the dynamical phase, and $%
\gamma_{n}( \boldsymbol{k})=\int_{\boldsymbol{k}_{0}}^{\boldsymbol{k}}\text{d%
}\boldsymbol{k}^{\prime}\cdot\left[\mathcal{R}_{\boldsymbol{k}^{\prime}}%
\right]_{n,n}$ is the Berry phase, where $\left[\mathcal{R}_{\boldsymbol{k}}%
\right]_{n,m}=\left\langle u_{n, \boldsymbol{k}}\right\vert i \frac{\partial%
}{\partial\boldsymbol{k}} \left\vert u_{m,\boldsymbol{k}}\right\rangle$ is
the Berry connection between bands $n$ and $m$.

A finite electric field causes non-adiabaticity, where the wavefunction in
general is a superposition of multiple bands: $\left\vert{u}%
\right\rangle=\sum_{n}\eta_{n}\left(t\right)e^{i\gamma_{n} (\boldsymbol{k}%
)}\left\vert{u}_{n, \boldsymbol{k}}\right\rangle$. The electron velocity
then has two parts $\dot{\boldsymbol{x}} \equiv \left\langle u\left\vert
\frac{\partial\mathcal{H}\left(\boldsymbol{k}\right)}{\hbar \partial%
\boldsymbol{k}}\right\vert u\right\rangle =\boldsymbol{v}_{b}+\boldsymbol{v}%
_{h}$,
\begin{equation}
\boldsymbol{v}_{b}=\sum_{n}\left\vert\eta_{n}\right\vert^{2}\frac{\partial
\varepsilon_{n,\boldsymbol{k}}}{\hbar \partial\boldsymbol{k}},  \label{vb-1}
\end{equation}
\begin{align}
\boldsymbol{v}_{h} & =\sum_{m,n}\eta^{*}_{m} \eta_{n} e^{i(\gamma_{n}(%
\boldsymbol{k})-\gamma_{m}(\boldsymbol{k}))} \left\langle u_{m}\left\vert
\frac{\partial\mathcal{H}}{\hbar \partial \boldsymbol{k}}\right\vert
u_{n}\right\rangle +\text{c.c.}  \label{vh-1}
\end{align}
$\boldsymbol{v}_{b}$ is the normal part from band dispersions. $\boldsymbol{v%
}_{h}$ is the anomalous velocity from inter-band coherences, which is our
focus here. The band amplitudes $\eta_{n}$ are subject to the equation of
motion,
\begin{align}
\bar{\mathcal{H}} \boldsymbol{\eta}=i\hbar\dot{\boldsymbol{\eta}},
\label{effSCE-1}
\end{align}
where $\boldsymbol{\eta}$ is a column vector whose entries are $\eta_{n}$.
For the example of massive Dirac fermion with only two bands ($n=c,v$),
\begin{equation}
\bar{\mathcal{H}}=\left(%
\begin{array}{cc}
\varepsilon_{c,\boldsymbol{k}} & \left[\bar{\mathcal{R}}_{\boldsymbol{k}}%
\right]_{c,v}\cdot\left(e\boldsymbol{E}\right) \\
\left[\bar{\mathcal{R}}_{\boldsymbol{k}}\right]_{c,v}^{*} \cdot\left(e%
\boldsymbol{E}\right) & \varepsilon_{v,\boldsymbol{k}}%
\end{array}%
\right),  \label{effvH-1}
\end{equation}
where $\left[\bar{\mathcal{R}}_{\boldsymbol{k}}\right]_{c,v}
=e^{-i\gamma_{c}\left(\boldsymbol{k}\right)}\left[\mathcal{R}_{\boldsymbol{k}%
}\right]_{c,v}e^{i\gamma_{v}\left(\boldsymbol{k}\right)}$. $\bar{\mathcal{H}}
$ is effectively a Hamiltonian in the moving frame of the carrier, whose $k$%
-space motion by the electric field gives rise to the coherent hybridization
between the Bloch bands, manifested as the off-diagonal term of Eq.~(\ref%
{effvH-1}). The solution of $\boldsymbol{\eta}$, formally expressed by
\begin{align}
\boldsymbol{\eta}\left(t\right)=\hat{T}\exp\left\{ -i\int_{0}^{t}\text{d}\tau%
\bar{\mathcal{H}}\left(\boldsymbol{k}\left(\tau\right),\boldsymbol{E}%
\left(\tau\right)\right) \right\} \boldsymbol{\eta}\left(0\right),
\label{sol-effSCE-1}
\end{align}
with $\hat{T}$ the time ordering operator, necessarily contains
non-perturbative effects of electric field to all orders.

Keeping the electric field effect to the leading order only, the above
solution for an electron initially in band $v$ reduces to~\cite{Boehm03book}
\begin{equation}
\left\vert u\right\rangle =e^{id_{v} (t)}\left( e^{i\gamma_{v}\left(%
\boldsymbol{k}\right)} \left\vert u_{v,\boldsymbol{k}}\right\rangle +r\left(%
\boldsymbol{k},\boldsymbol{E}\right) e^{i\gamma_{c}\left(\boldsymbol{k}%
\right)} \left\vert u_{c,\boldsymbol{k}}\right\rangle \right) ,  \notag
\end{equation}
where,
\begin{equation}
r\left(\boldsymbol{k},\boldsymbol{E}\right)= \frac{ - \left[\bar{\mathcal{R}}%
_{\boldsymbol{k}}\right]_{c,v}\cdot\left(e\boldsymbol{E}\right)} {%
\varepsilon_{c,\boldsymbol{k}} -\varepsilon_{v,\boldsymbol{k}}},
\label{coef-1stod-1}
\end{equation}
a dimensionless quantity measuring the ratio between the field strength to
the gap size. The anomalous velocity from the interband coherence then
becomes,
\begin{eqnarray}
\boldsymbol{v}_{h} &=& r(\boldsymbol{k}, \boldsymbol{E}) e^{i(\gamma_{c}(%
\boldsymbol{k})-\gamma_{v}(\boldsymbol{k}))} \left\langle {u}_{v}\left\vert
\frac{\partial\mathcal{H}\left(\boldsymbol{k}\right)}{\partial\hbar%
\boldsymbol{k}}\right\vert {u}_{c}\right\rangle +\text{c.c.}  \notag \\
&=&-\frac{e}{\hbar}\boldsymbol{E}\times\boldsymbol{\Omega}_v \left(%
\boldsymbol{k}\right).  \label{vh-2}
\end{eqnarray}
Here by retaining the leading order band hybridization of the moving
electron driven by electric field (Eq.(\ref{effvH-1})), the intrinsic Hall
effect from the band Berry curvature $\boldsymbol{\Omega}_v =\boldsymbol{%
\nabla}\times\left[\mathcal{R}_{\boldsymbol{k}}\right]_{v,v}$ is recovered.
Importantly, concerning the Hall current for an ensemble of
electrons,
\begin{equation}
\boldsymbol{J}^{\text{H}}=-e\int d\boldsymbol{k}f\left(\boldsymbol{k}\right)
\boldsymbol{v}_{h},  \label{sclJ1}
\end{equation}
in this \textit{first order} adiabatic approximation, $\boldsymbol{v}_{h}$
is purely a function of $\boldsymbol{k}$ without other time dependence, so
the Hall conductance is completely determined by the instantaneous Fermi
distribution $f(\boldsymbol{k})$. A steady state dc Hall current is obtained
when $f\left(\boldsymbol{k}\right)$ is stabilized by momentum relaxation.

\begin{figure}[h]
 \includegraphics[width=8.6cm,height=5.8cm]{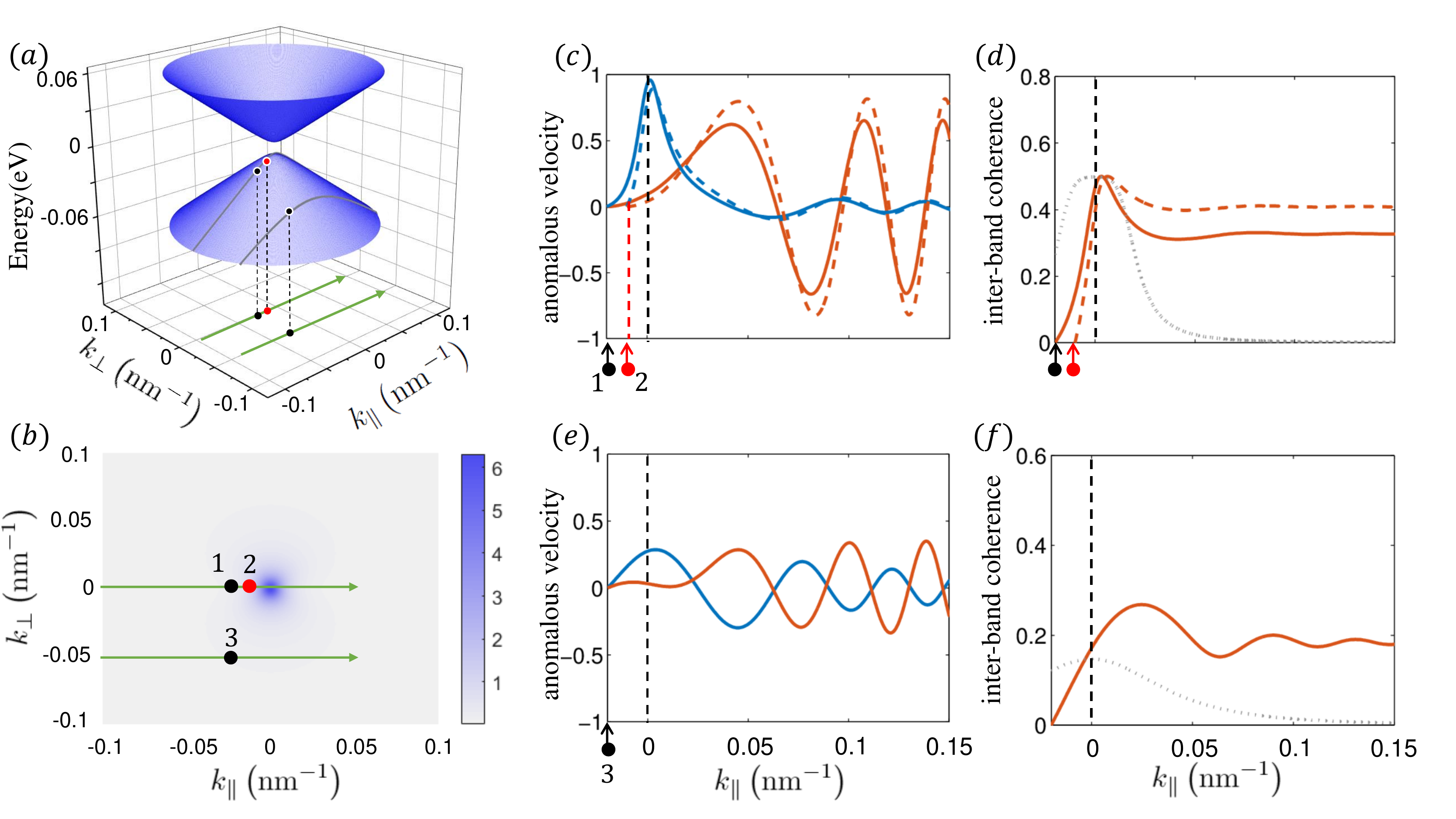}
\caption{Coherent non-adiabatic dynamics. (a) Dirac cone with gap $\Delta=0.01$eV. (b) $r$ as function of $\boldsymbol{k}$ at a field of $E=1$V/$\mu$m (c.f. Eq.(\ref{coef-1stod-1})). (c) Anomalous velocity components perpendicular ($v_{h}^{\perp}$, in orange) and parallel ($v_{h}^{\parallel}$, in blue) to the field (Eq.(\ref{vh-1})). Solid and dashed trajectories are for the initial $\boldsymbol{k}_{0}$ labeled as 1 and 2 respectively in (b). (d) The corresponding inter-band coherence $\left\vert\eta_{v}^{*}\eta_{c}\right\vert$. (e) and (f) Same plots for the initial $\boldsymbol{k}_{0}$ labeled as 3 in (b).
$\boldsymbol{k}$ in unit of $\text{{\AA}}^{-1}$ and $\boldsymbol{v}_{h}$ in unit of $v_{D}$.
The gray-dotted lines in (d) and (f) are steady-state values of $\left\vert\eta_{c}^{*}\eta_{v}\right\vert$ from incoherent dynamics for comparison.}
\label{schematics}
\end{figure}

In contrast, in the general non-adiabatic dynamics at large $r$, when the
electric field is retained to all orders in Eq.~(\ref{sol-effSCE-1}), $%
\boldsymbol{v}_{h}$ is not a pure function of $\boldsymbol{k}$, but has
additional time dependence determined by the evolution history. Fig.~\ref{schematics} shows
examples of numerically calculated exact evolutions for a massive Dirac cone with the mass $\Delta =0.01\mathrm{eV}$ in an electric field of $1\mathrm{V/\mu m}$, which corresponds to $r\approx 6$ at the Dirac point. $\boldsymbol{v}_{h}(t)$ are oscillating
functions of time, with amplitudes and phases dependent on the initial $%
\boldsymbol{k}_{0}$.
Interestingly, when $\boldsymbol{k}$ evolves far away
from the Dirac point, the component parallel to the electric field ${v}_{h}^{\parallel}$ damps
out, while the perpendicular component ${v}_{h}^{\perp }$ reaches
a constant amplitude, corresponding to an \textit{ac} Hall velocity.
This is evident by applying Eq.~(\ref{vh-1}) to the Dirac Hamiltonian, which leads to%
\begin{eqnarray}
{v}_{h}^{\perp } &=&\frac{2v_{D}}{k}\left( k_{\parallel }%
\mathrm{\mathrm{Im}}Z-\frac{k_{\perp }\Delta }{\varepsilon _{v,\boldsymbol{k}%
}-\varepsilon _{c,\boldsymbol{k}}}\mathrm{Re}Z\right) , \\
{v}_{h}^{\parallel } &=&\frac{2v_{D}}{k}\left( -k_{\perp }\mathrm{Im}Z-\frac{k_{\parallel }\Delta }{\varepsilon _{v,\boldsymbol{k}%
}-\varepsilon _{c,\boldsymbol{k}}}\mathrm{Re}Z\right) ,  \notag
\end{eqnarray}%
where $Z=e^{i(\gamma _{c}(\boldsymbol{k})-\gamma _{v}(\boldsymbol{k}))}\eta
_{v}^{\ast }\eta _{c}$, $v_{D}$ is the
Dirac velocity and $k_{\parallel }$ ($k_{\perp }$) denotes the component of $\boldsymbol{k}$ parallel (perpendicular) to $\boldsymbol{E}$.
On the other hand, this oscillation, and the history dependence, apparently
raise the issue of how steady state \textit{dc} current can be reached in
the non-adiabatic regime.

In reality, interband coherence can not be retained in arbitrary form in the
presence of decoherence due to the environment. To account for such effect,
we introduce a stochastic noise to the evolution.
Writing $\bar{\mathcal{H}} \equiv \boldsymbol{h}(\boldsymbol{k}) \cdot \boldsymbol{\sigma}$, $\boldsymbol{\sigma}$ the vector of Pauli matrices, this Hamiltonian-like term  in Eq.~(\ref{effSCE-1}) is now replaced by $ \left[ \boldsymbol{h} + F(t) \hat{\boldsymbol{l}} \right] \cdot\boldsymbol{\sigma}$, where $\hat{\boldsymbol{l}}$ a unit vector specified by polar angle $\theta$ and azimuthal angle $\phi$ with respect to $\boldsymbol{h}$.
The stochastic force has zero mean $\left\langle{F(t)}\right\rangle_s=0$ and noise
correlation $\left\langle{F(t)F(t^{\prime })}\right\rangle_s=\int\frac{\text{%
d}\omega}{2 \pi}\left[n(\omega){e}^{i\omega\left(t-t^{\prime}\right)}+{e}%
^{-i\omega\left(t-t^{\prime}\right)} \left(n(\omega)+1\right)\right]%
J\left(\omega\right)$ specified by the spectral density $J\left(\omega%
\right) $. $\left\langle\cdots\right\rangle_s$ denotes the statistical
average over noise trajectories, and $n(\omega)=1/\left[e^{\omega/k_{B}T}-1%
\right]$. Instead of the coherent wavefunction ${\boldsymbol{\eta}}$, we
turn to the noise-averaged density matrix $\rho(t)= \langle {\boldsymbol{\eta%
}}\left(t\right) {\boldsymbol{\eta}}^{\dagger}\left(t\right) \rangle_s$.
Under the Born-Markov approximation for the noise spectrum,~\cite%
{Shnirman02147,Leggett871,Breuer02book}
\begin{equation}
\dot{\rho}=\frac{i}{\hbar}\left[\rho,\bar{\mathcal{H}}\right]-\frac{1}{2
\hbar^2}\left[\hat{\boldsymbol{n}} \cdot \boldsymbol{\sigma},\Gamma_{-} \rho
- \rho \Gamma_{+}\right].  \label{densitymatrix}
\end{equation}
$\Gamma_{\pm}= J\left(\bar{\varepsilon}_{g}\right) \sin\theta \left[ {e^{\mp
i\phi}} \boldsymbol{z}_\pm \boldsymbol{z}_\mp ^{\dagger} +n(\bar{\varepsilon}%
_{g})\left(e^{i\phi}\boldsymbol{z}_- \boldsymbol{z}_+^{\dagger} + h.c.
\right) \right] + J\left(0\right) (2n(0)+1) \cos\theta (\boldsymbol{z}_+
\boldsymbol{z}_+^{\dagger} - \boldsymbol{z}_- \boldsymbol{z}_-^{\dagger}) $.
$\boldsymbol{z}_{\pm}$ denote eigenvectors of $\bar{\mathcal{H}}$,
\begin{equation}
\bar{\mathcal{H}}(\boldsymbol{k, E}) \boldsymbol{z}_{\pm}=\pm\left(\bar{%
\varepsilon}_{g}/2\right)\boldsymbol{z}_{\pm},  \label{eigenvector}
\end{equation}
which are coherent hybridizations of bands $c$ and $v$, and
\begin{equation}
\bar{\varepsilon}_{g} = \left(\varepsilon_{c,\boldsymbol{k}} -
\varepsilon_{v,\boldsymbol{k}} \right) \sqrt{1 + 4 \left\vert r(\boldsymbol{k%
},\boldsymbol{E}) \right\vert^{2}} .  \label{eg-hyb0}
\end{equation}

The steady state solution of Eq.~(\ref{densitymatrix}) is
\begin{equation}
\rho = p_+ \boldsymbol{z}_+ \boldsymbol{z}_+^{\dagger} + p_- \boldsymbol{z}%
_- \boldsymbol{z}_-^{\dagger},
\end{equation}
where
\begin{equation}
p_+=\frac{{n}(\bar{\varepsilon}_{g})}{1+2{n}(\bar{\varepsilon}_{g})}, ~~~
p_-=\frac{1+{n}(\bar{\varepsilon}_{g})}{1+2{n}(\bar{\varepsilon}_{g})}.
\label{prob}
\end{equation}
The joint actions of electric field and the noise thus favor certain forms
of coherent interband hybridization, i.e. eigenstates of the effective
Hamiltonian $\bar{\mathcal{H}}$. At low temperature ${n}(\bar{\varepsilon}%
_{g}) \rightarrow 0$, $p_+ \rightarrow 0, p_- \rightarrow 1$, the $%
\boldsymbol{z}_{-}$ state is favored. While the treatment of decoherence
here is oversimplified, the above message can be rather general.

The decoherence process washes away the coherence between $\boldsymbol{z}%
_{+} $ and $\boldsymbol{z}_{-}$, each of which however has retained the
interband coherence induced by the electric field in a non-perturbation
manner. Plugging ${\boldsymbol{\eta}}=\boldsymbol{z}_{-}$ into Eqs.~(\ref%
{vb-1}) and (\ref{vh-1}), the velocities that include electric field effects
to all orders are obtained
\begin{eqnarray}
\boldsymbol{v}_b&=& \frac{\partial\varepsilon_{v, \boldsymbol{k}}}{\hbar
\partial\boldsymbol{k}} \frac{1}{\sqrt{1+4 \left\vert r(\boldsymbol{k},%
\boldsymbol{E}) \right\vert^{2} }} , \\
\boldsymbol{v}_h&=& -\frac{e}{\hbar}\boldsymbol{E}\times\boldsymbol{\Omega}%
_{v}\left(\boldsymbol{k}\right) \frac{1}{\sqrt{1+4 \left\vert r(\boldsymbol{k%
},\boldsymbol{E}) \right\vert^{2}}}.  \label{twobnd-eqx}
\end{eqnarray}
The velocities associated with the $\boldsymbol{z}_{+}$ state has the same
form, with the index $v$ replaced by $c$. In comparison to Eq.~(\ref{vh-2}),
the non-adiabatic effects of electric field are manifested simply through
the prefactor $1/\sqrt{1+ 4 \left\vert r(\boldsymbol{k},\boldsymbol{E})
\right\vert^{2} }$, and the anomalous velocity in the adiabatic limit is
reproduced at $r\ll1$. Remarkably, in the non-adiabatic anomalous velocity,
its orthogonality to the applied electric field is retained, signifying the
intrinsic role of the Berry curvature in generating the anomalous motion.
Besides, Eq.~(\ref{twobnd-eqx}) contains only the odd orders of the electric
field. This agrees with the intuition that reversing the direction of the
electric field reverses the direction of the induced carrier's motion.
The dotted curves in Fig.~\ref{schematics} (d) and (f) plot the underlying interband coherence $\left\vert\eta_{c}^{*}\eta_{v}\right\vert$, which is peaked at the Dirac point, i.e. the Berry curvature hot spot.

In Eq.~(\ref{prob}), the eigenvalue of $\bar{\mathcal{H}}(\boldsymbol{k})$
is apparently playing the role of renormalised energy in determining the
probability distribution of electron at given momentum $\boldsymbol{k}$.
This, combined with momentum relaxation, further allows us to discuss a
steady state distribution $f(\boldsymbol{k})$. The Hall current of an
electron ensemble then follows from Eq.~(\ref{sclJ1}) and (\ref{twobnd-eqx}). From Eq.~(\ref
{twobnd-eqx}), we recover the concerned feature that $\boldsymbol{v}_{h}$ is
a function of $\boldsymbol{k}$ only, and a steady Hall current that is
completely determined by $f(\boldsymbol{k})$ again. Specifically, for a
massive Dirac cone at its charge neutrality, the electrons occupy the branch
$\boldsymbol{z}_{-}$ at all $\boldsymbol{k}$, with a thermal
excitation gap $\Delta \sqrt{1+\hbar ^{2}v_{D}^{2}\left\vert e\boldsymbol{E}%
\right\vert ^{2}/\Delta ^{4}}$. At the low
temperature limit, Eq.~(\ref{sclJ1}) becomes,
\begin{align}
\boldsymbol{J}^{\text{H}}=\frac{e^{2}}{\hbar}\int d\boldsymbol{k}
\frac{\boldsymbol{E}\times\boldsymbol{\Omega}_{v}\left(\boldsymbol{k}\right)}
{\sqrt{1+4 \left\vert r(\boldsymbol{k%
},\boldsymbol{E}) \right\vert^{2}}},
\end{align} and
the Hall conductance up to third order response reads,
\begin{equation}
\sigma_{\text{H}}\approx \frac{\text{sgn}(\Delta )}{2}\frac{e^{2}}{h}\left[ 1-%
\frac{27}{35}\frac{\hbar ^{2}v_{D}^{2}\left( eE\right) ^{2}}{\Delta^{4}}\right]. \notag
\end{equation}%

\begin{figure}[h]
\includegraphics[width=8.5cm,height=3.8cm]{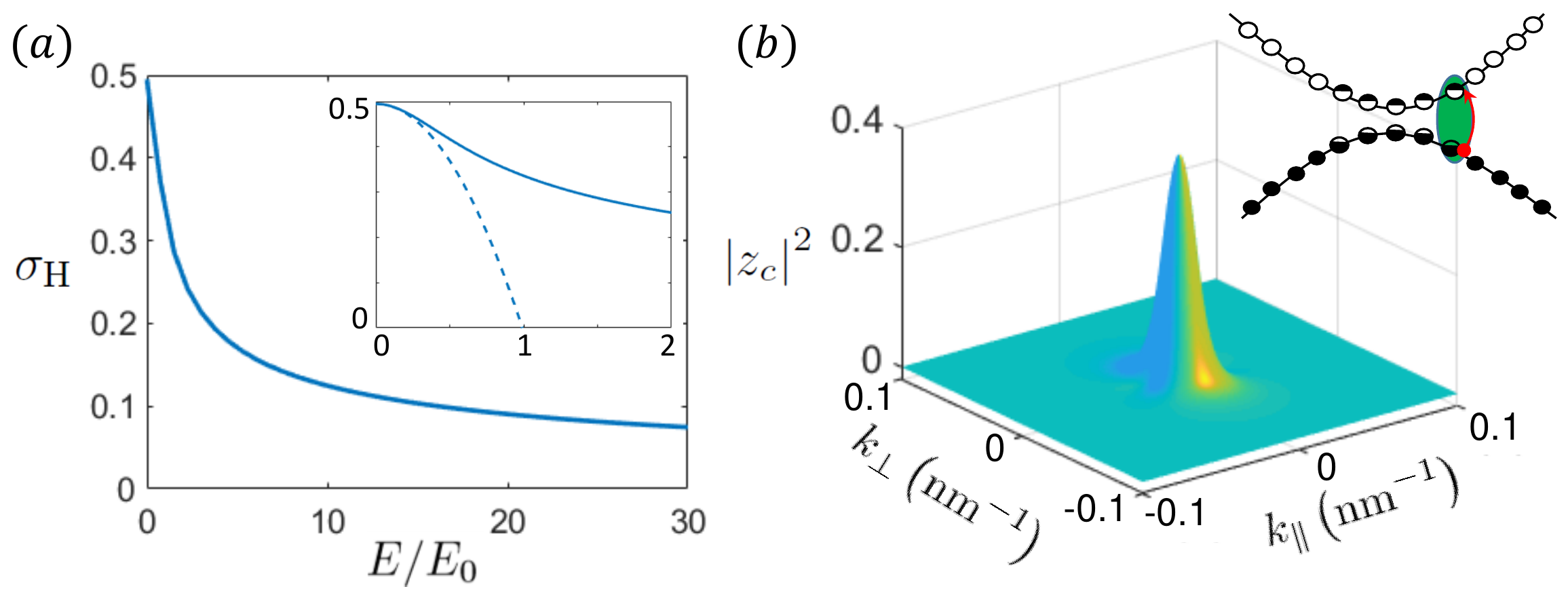}
\caption{Incoherent non-adiabatic Hall effect in a gapped Dirac cone. (a) The non-perturbative dc Hall conductance as function of electric field, in unit of $e^{2}/\hbar$. Dashed curve in the inset keeps up to the third order response. (b) Electron-hole pair density as a function of $\boldsymbol{k}$, in the steady state at $E=2E_0$.}
\label{Hc}
\end{figure}

We show in Fig.~\ref{Hc}(a) the field-dependence of the Hall
conductance, in comparison with the half-quantized value $\sigma_{\text{H}} (E=0)$ from the linear
response part.~\cite{Xiao07236809} Define a critical field $E_0$
by $\left\vert\left[\mathcal{R}_{\boldsymbol{k}=0}\right]_{c,v}\right\vert{%
eE_0}=\Delta$, i.e. making $r=1$ at the Dirac point, the deviation $\left\vert\sigma_{\text{H}}-\sigma_{\text{H}}(0)\right\vert/\sigma_{\text{H}}$ readily grows to 10\% when ${E/E_0}\approx0.3$. Keeping up to third order
response has limited improve (see the inset of Fig.~\ref{Hc}(a)), showing the
non-perturbative nature in this field range marked by $E_0$. For $%
\Delta=0.01 $eV, $E_0=0.167$V/$\mu$m.

\section{Discussion}  

The above results establish a general connection
between interband coherence and intrinsic Hall effect in both the coherent
and incoherent dynamics, from the adiabatic to the non-adiabatic regime.
This can help to elucidate the current carrying mechanism of the pseudospin Hall
effect in an insulator.~\cite{Gorbachev14448,Sui151027,Shimazaki151032,Liu2001.08401}
 For the Hall current described by Eq.~(\ref{sclJ1}),
the underlying steady state can be written as
\begin{equation}
\left\vert \Psi \right\rangle = \prod_{\boldsymbol{k}} (z_{v} + z_{c}
a^{\dagger}_{c,\boldsymbol{k}} a_{v,\boldsymbol{k}}) \left\vert \text{vac}%
\right\rangle,  \notag
\end{equation}
where $\left\vert \text{vac}\right\rangle$ denotes the equilibrium Fermi sea
with filled (empty) valence (conduction) band. $z_{v}$ and $z_{c}$ are
entries of eigenvector $\boldsymbol{z}_{-}$ in Eq.~(\ref{eigenvector}) that
depend on both $\boldsymbol{k}$ and $\boldsymbol{E}$. In the
non-perturbative regime, the electric field creates electron-hole pairs with
sizeable probability on top of the Fermi sea (c.f. Fig.~\ref{Hc}(b)). This is in contrast to the impression conveyed by the linear response picture where the
Fermi sea is considered unperturbed by the infinitesimal field (effect on
band filling $\propto E^2$ neglected).
Instead of a response purely by the equilibrium Fermi sea, the Hall current, expressed as
\begin{align}
\boldsymbol{J}^{\text{H}} = \sum_{\boldsymbol{k}} z_{v}^{*}z_{c}
\left\langle \text{vac}\left\vert \hat{\boldsymbol{J}}a^{\dagger}_{c,%
\boldsymbol{k}} a_{v,\boldsymbol{k}} \right\vert \text{vac}\right\rangle +%
\text{c.c.}  \notag
\end{align}
where $\hat{\boldsymbol{J}}$ is the current operator, comes from the
interference of the electron-hole pair excitations $%
a^{\dagger}_{c,\boldsymbol{k}} a_{v,\boldsymbol{k}} \left\vert\text{vac}%
\right\rangle$ with the Fermi sea background $\left\vert\text{vac}\right\rangle$.
The bulk pseudospin Hall current
can thus be sustained without conducting states inside the gap. In the
adiabatic limit $r \rightarrow 0$, the pair-excitation probability becomes
infinitesimally small, and the above picture properly reduces to the linear
response one.

\begin{acknowledgements}
The work is supported by the Research Grants Council of Hong Kong (Grants No. HKU17306819 and No. C7036-17W), and the University of Hong Kong (Seed Funding for Strategic Interdisciplinary Research).
\end{acknowledgements}

\end{document}